\documentclass[11pt,twoside
]{article}

\usepackage{baaa2009}
\usepackage{graphicx}
\usepackage{subfigure}
\usepackage{psfrag}
\usepackage{amssymb}
\usepackage[spanish,activeacute,english]{babel}
\usepackage[latin1]{inputenc}
\usepackage[T1]{fontenc} 
\usepackage{ae,aecompl} 
\usepackage{latexsym}
\usepackage{verbatim}
\usepackage{amsmath}
\usepackage{stmaryrd}
\usepackage{amsfonts}
\usepackage{amssymb}
\usepackage{wasysym}
\usepackage[colorlinks=true,dvips]{hyperref}

\begin{document}
\myselectenglish
\vskip 1.0cm
\markboth{ Smith Castelli et al. }%
{cE galaxies in Antlia}

\pagestyle{myheadings}
\vspace*{0.5cm}
\noindent PRESENTACIÓN ORAL
\vskip 0.3cm
\title{Two confirmed compact elliptical galaxies in the Antlia cluster}


\author{Anal\'ia V. Smith Castelli$^{1,2,3}$, Favio R. Faifer$^{1,2,3}$, Lilia
P. Bassino$^{1,2,3}$, Gisela A. Romero$^{1,4}$ Sergio A. Cellone$^{1,2,3}$, Tom Richtler$^{5}$}

\affil{%
  (1) Facultad de Ciencias Astronómicas y Geofísicas - UNLP, Argentina\\
  (2) CONICET, Argentina\\
  (3) Instituto de Astrof\'isica La Plata (CCT-La Plata, CONICET), Argentina\\
  (4) Universidad de Valpara\'iso, Chile\\
  (5) Universidad de Concepci\'on, Chile\\
}

\begin{abstract} We confirm the existence of two compact elliptical (cE)
galaxies in the central region of the Antlia cluster through MAGELLAN-MIKE 
and GEMINI-GMOS spectra. Only about a dozen galaxies of this rare type are 
known today up to a distance of 100 Mpc. With this finding, Antlia becomes 
the nearest galaxy cluster harbouring more than one cE galaxy among its 
galaxy population. One of these galaxies shows evidence of interaction with 
one of the giant ellipticals that dominate the central region of the cluster.
\end{abstract}

\begin{resumen}
  A trav\'es de espectros obtenidos con MAGALLANES-MIKE y GEMINI-GMOS, 
  confirmamos la existencia de dos galaxias de tipo compacta elliptica (cE)
  en la regi\'on central del c\'umulo de Antlia. S\'olo se conocen alrededor
  de una docena de galaxias de este raro tipo hasta la fecha, y hasta una 
  distancia de
  100 Mpc. Con este hallazgo Antlia se convierte en el c\'umulo de galaxias
  m\'as cercano en presentar m\'as de una cE entre sus galaxias 
  miembro. Una de estas galaxias muestra evidencia de interacci\'on con una
  de las galaxias el\'ipticas gigantes que dominan la regi\'on central del 
  c\'umulo.
\end{resumen}

\section{Introduction}
\subsection{The Antlia cluster}

The Antlia cluster is the third nearest galaxy cluster after Virgo and Fornax.
Dirsch et al. (2003) calculated a distance modulus of $(m-M)=32.73$, 
which translates into a distance of 35.2 Mpc. Sandage (1975) described Antlia 
as a {\it beautiful small group dominated by the two equally bright E galaxies 
NGC\,3258 and NGC\,3268.} He noticed that {\it its galaxy population consists 
of a few spiral galaxies among its Es and S0s.} Ferguson \& Sandage (1990, 
hereafter FS90) identified 375 galaxies within a projected area of $\sim 8$ 
Mpc$^2$ that are listed in the FS90 Antlia Group Catalogue. Only 6\% of 
these galaxies had redshift information at that moment. FS90 confirmed 
the elongated structure of Antlia, already noticed by Hopp \& Materne (1985). 
They also estimated that its central galaxy density is 1.7 times higher than in 
Virgo and 1.4 times higher than in Fornax.  

Antlia's galaxy population was poorly studied until we began our Antlia Cluster
Project. This is a long-term project aimed at performing a large-scale 
photometric and spectroscopic study of the galaxy content of the cluster. 
Our previous results refer to the globular cluster (GC) systems of NGC\,3258 
and NGC\,3268 (Dirsch et al. 2003, Bassino et al. 2008), and to the 
early-type galaxy population in the central region of the cluster (Smith 
Castelli et al. 2008a,b, hereafter Paper I and Paper II, respectively; 
Smith Castelli 2008).

\subsection{Compact elliptical galaxies}

Compact elliptical (cE) galaxies constitute a very rare type of objects as 
only about a dozen have been identified up to a distance of 100 Mpc (e.g. 
Chilingarian et al. 2007, 2009; Price et al. 2009). 
The prototype of this class is M32 (but see Graham 2002) and their main 
characteristics are a high central and effective surface brightness for their
luminosities, and a high degree of compactness. Most of the known examples
are placed close in projection to giant galaxies. FS90 classified 11 objects
as possible cE galaxies in the Antlia cluster. Two of them are members of
the cluster but their morphologies do not match that of cE systems.

In this contribution we present two newly confirmed FS90 cE galaxies in 
Antlia (namely, FS90\,110 and FS90\,192). Each of them is close in projection 
to one of the giant ellipticals NGC\,3258 and NGC\,3268 (Fig.\,1). A 
photometric analysis was presented in Paper II. In that work, both 
galaxies were considered as firm candidates to be genuine cEs due to their 
photometric characteristics, similar to those of confirmed cEs.  
It is interesting to find new members of this class in the nearby Universe as 
it has been proposed that cE galaxies are the low-mass counterparts of giant 
ellipticals (Kormendy et al. 2009).    

\section{Observations}

MAGELLAN-MIKE echelle spectra of FS90\,110 and FS90\,192 were obtained
at the CLAY telescope of Las Campanas Observatory in 2009 March 27 and 28 
(program ID: LCO-CNTAC09A\_042). Slits of 1 x 5 arcsec and binning 2 x 3 were 
used. The spectral coverage at the red side of the echelle spectrograph was 
4900-10000 \AA, with a resolution (fwhm) of $\sim 0.35$\,\AA. For FS90\,110, 
two spectra of 900 sec and one of 2400 sec were obtained, and only one of 2400 
sec for FS90\,192. The reduction was performed using a combination of the 
\textsf{imred.ccdred.echelle} and \textsf{mtools} packages within IRAF.    

\begin{figure}[!h]
  \centering
    \includegraphics[width=.4\textwidth,angle=-90]{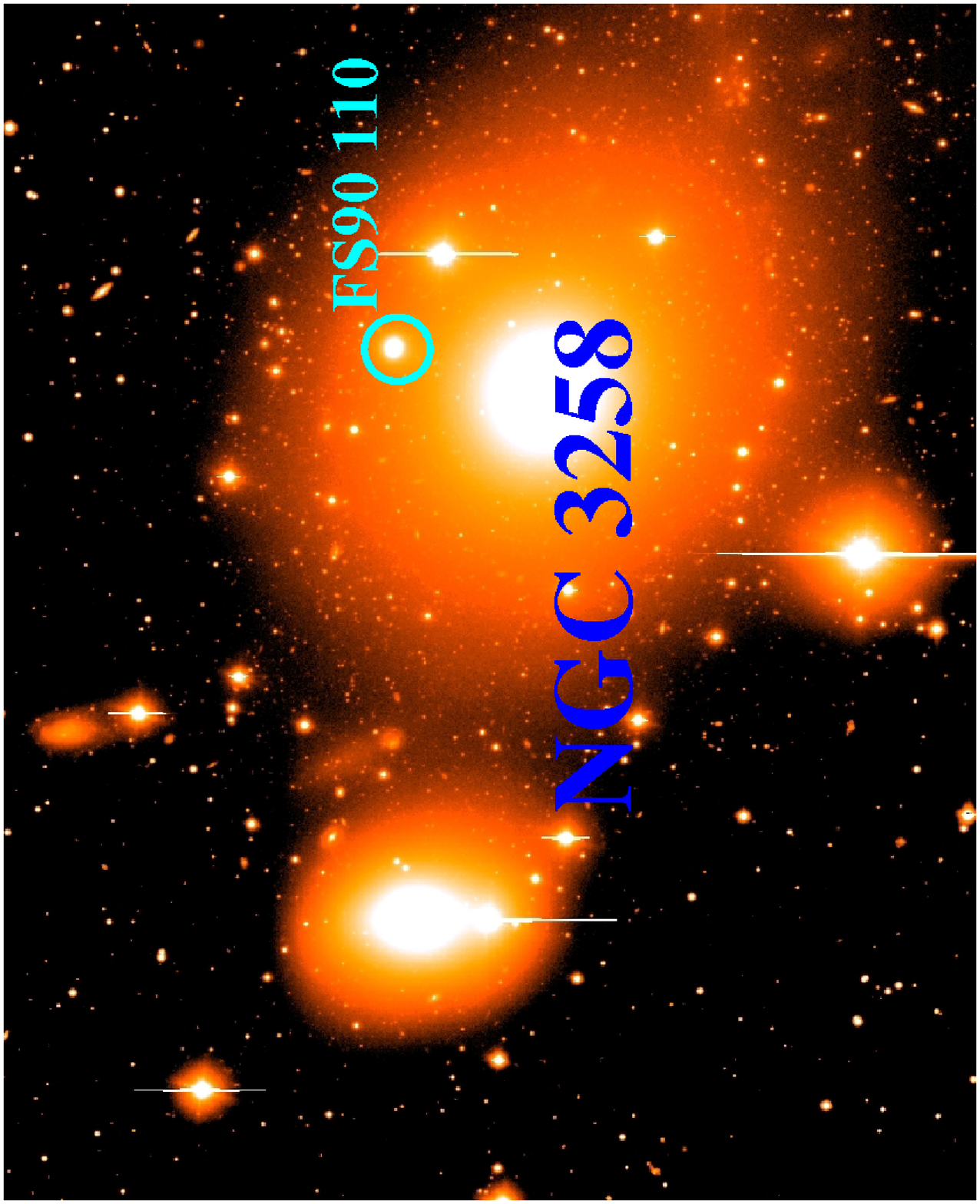}~\hfill%
    \includegraphics[width=.4\textwidth,angle=-90]{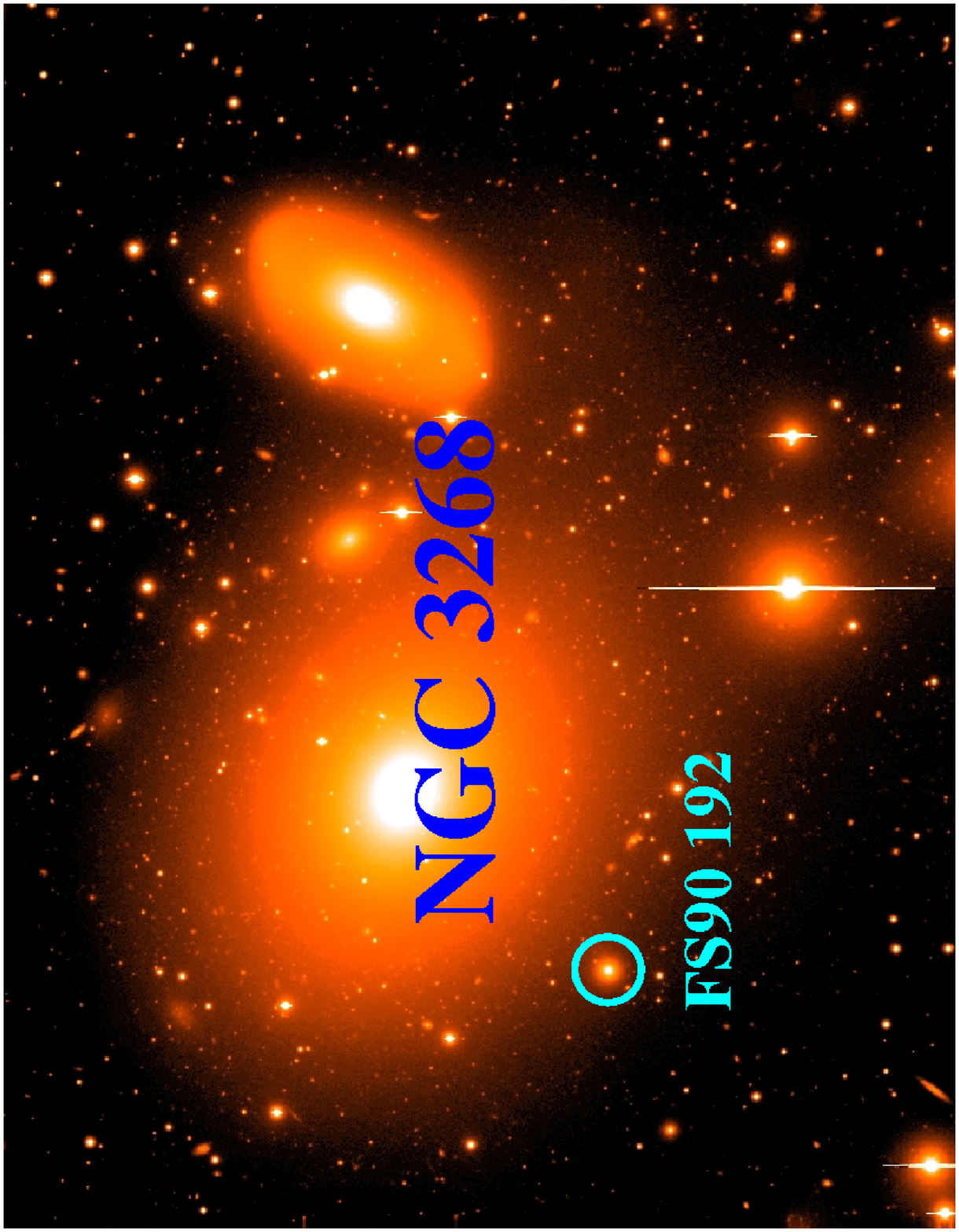}~\hfill%
  \caption{VLT-FORS1 images of NGC\,3258 and FS90\,110 (left), and NGC\,3268 
           and FS90\,192 (right) (Bassino et al. 2008). North is up, and east 
           to the left.}
  \label{fig1}
\end{figure}

We have also obtained GEMINI-GMOS multi-object spectra for three fields placed 
in the central region of Antlia. FS90\,110 was located in one of these fields.
The data were taken during January and March 2009 at the Gemini-South 
Observatory (program ID: GS-2009A-Q-25, PI: L. Bassino, 8 hours in Band I). We 
used a slit width of 1 arcsec, and the B600\_G5303 grating blazed at 5000\,\AA
 with three different central wavelengths (5000, 5050 and 5100\,\AA) in order 
to fill in the CCD gaps. The wavelength coverage is 3500 - 7200\,\AA, 
depending on the positions of the slits, and the resolution (fwhm) is 
$\sim 4.6$\,\AA. The total on source integration time was 3.2 hours, 
comprising 5 exposures of 40 minutes each. Data reduction was performed in a 
standard manner using the \textsf{gemini.gmos} package within IRAF.

\section{Preliminary Results}

\begin{figure}[t]
  \centering
    \includegraphics[width=.55\textwidth]{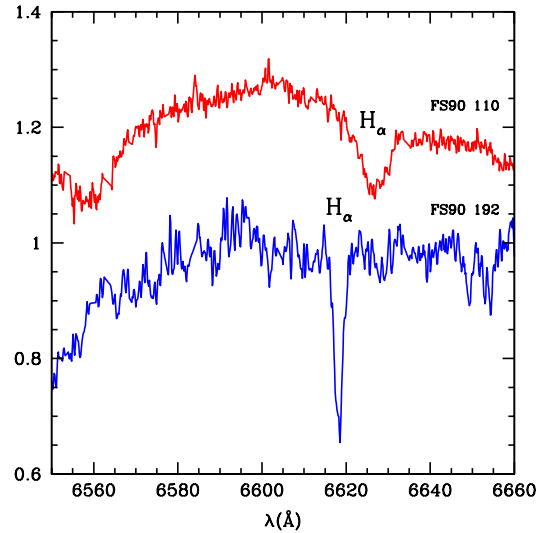}~\hfill%
  \caption{Echelle MAGELLAN-MIKE spectra of FS90\,110 (top) and FS90\,192
           (bottom) showing the position displayed by ${\rm H}_\alpha$ (6562.82 
	   \AA). The spectrum of FS90\,110 is shifted upwards in order to avoid 
	   superposition with that of FS90\,192. Both spectra are not flux 
	   calibrated. }
  \label{fig1}
\end{figure}

Through the identification of the more prominent absorption lines (i.e. 
${\rm H}_\alpha$, Na and MgI) on the red side of the echelle spectra, we have 
obtained preliminary radial velocities of $\sim$ 2900 km ${\rm s}^{-1}$ for 
FS90\,110, the neighbour of NGC\,3258, and $\sim$ 2500 km ${\rm s}^{-1}$ for 
FS90\,192, the cE companion of NGC\,3268 (Fig.\,2). We consider Antlia members 
those galaxies displaying radial velocities in the range 1200 -- 4200 km 
${\rm s}^{-1}$ (Paper I, Faifer et al. 2008). The value obtained for FS90\,110 
has been confirmed through our GMOS spectra and it is consistent with 
the radial velocity of NGC\,3258 ($\sim$ 2800 km $s^{-1}$, Paper I). It is 
interesting to recall that FS90\,110 shows a low surface brightness 
{\it bridge} that seems to link it to NGC\,3258 (Paper II). With these findings 
Antlia becomes the nearest galaxy cluster hosting more than one cE galaxy 
among its galaxy population. In addition, FS90\,110 is the only cE showing 
clear evidence of interaction with its bright companion. We will present a 
detailed spectroscopic and additional photometric analysis of these new cE 
galaxies in a forthcoming paper (Smith Castelli et al. in preparation).\\    

\agradecimientos

We are grateful to Nidia Morrell for teaching us how to reduce echelle 
spectra. This work was supported by grants from Consejo 
Nacional de Investigaciones Cient\'ificas y T\'ecnicas (CONICET).
G.A.R. was supported by ALMA FUND Grant 31070021.
                                                                      
\begin{referencias}
\vskip 1cm
         
\reference Bassino L. P., Richtler T., Dirsch B., 2008, \mnras, 386, 1145

\reference Chilingarian I., Cayatte V., Chemin L., Durret F., Lagan´a T. F., 
Adami C., Slezak E., 2007, A\&A, 466, L21

\reference Chilingarian I., Cayatte V., Revaz Y., Dodonov S., Durand D., Durret F., Micol A., Slezak E., 2009, Science, in press (arXiv0910.0293)

\reference Dirsch B., Richtler T., Geisler D., Forte J. C., Bassino L. P., Gieren W. P., 2003, \aj, 125, 1908

\reference Faifer F., Smith Castelli A.V., Bassino L.P., Richtler T., Cellone S.A., 2008, BAAA, 51, 227

\reference Ferguson H. C., Sandage A., 1990, \aj, 100, 1 (FS90)

\reference Graham A. W., 2002, \apj, 568, L13 (erratum 572, L121)

\reference Hopp U., Materne J., 1985, A\&AS, 61, 93

\reference Kormendy J., Fisher D.B., Cornell M.E., Bender R., 2009, ApJS, 182, 216

\reference Price J. et al., 2009, \mnras, 397, 1816

\reference Sandage A., 1975, \apj, 202, 563

\reference Smith Castelli A.V., 2008, Doctoral Thesis, UNLP, Argentina 

\reference Smith Castelli A.V., Bassino L.P., Richtler T., Cellone S.A., Aruta
C., Infante L., 2008a, \mnras, 386, 2311 (Paper I)

\reference Smith Castelli A.V., Faifer F.R., Richtler T., Bassino L.P., 2008b,
\mnras, 391, 685 (Paper II)
                                                                       
\end{referencias}

\end{document}